\documentclass[reprint,twocolumn,superscriptaddress,amsmath,amssymb,prb,longbibliography]{revtex4-1}

\pdfoutput=1 

\usepackage[english]{babel}
\usepackage{dcolumn}
\usepackage{epsfig}
\usepackage{graphics}
\usepackage[latin1]{inputenc} 
\usepackage[T1]{fontenc}
\usepackage{bm}
\usepackage{hyperref}
\newcommand{\ddst}{false}

\begin{document}

\title{Topological Control on Silicates Dissolution Kinetics}

\author{Isabella Pignatelli}
  \affiliation{Laboratory for the Chemistry of Construction Materials (LC$^2$), Department of Civil and Environmental Engineering, University of California, Los Angeles, CA, USA}
\author{Aditya Kumar}
  \affiliation{Laboratory for the Chemistry of Construction Materials (LC$^2$), Department of Civil and Environmental Engineering, University of California, Los Angeles, CA, USA}
\author{Mathieu Bauchy}
  \email[Contact: ]{bauchy@ucla.edu}
  \homepage[\\Homepage: ]{http://mathieu.bauchy.com}
  \affiliation{Physics of AmoRphous and Inorganic Solids Laboratory (PARISlab), Department of Civil and Environmental Engineering, University of California, Los Angeles, CA, USA}
\author{Gaurav Sant}
  \email[Contact: ]{gsant@ucla.edu}
  \homepage[\\Homepage: ]{http://www.lcc-ucla.edu}
  \affiliation{Laboratory for the Chemistry of Construction Materials (LC$^2$), Department of Civil and Environmental Engineering, University of California, Los Angeles, CA, USA}
  \affiliation{California Nanosystems Institute (CNSI), University of California, Los Angeles, CA, USA}

\date{\today}


\begin{abstract}
Like many others, silicate solids dissolve when placed in contact with water. In a given aqueous environment, the dissolution rate depends highly on the composition and the structure of the solid, and can span several orders of magnitude. Although the kinetics of dissolution depends on the complexities of both the dissolving solid and the solvent, a clear understanding of which critical structural descriptors of the solid control its dissolution rate is lacking. Through pioneering dissolution experiments and atomistic simulations, we correlate the dissolution rates -- ranging over four orders of magnitude -- of a selection of silicate glasses and crystals, to the number of chemical topological constraints acting between the atoms of the dissolving solid. The number of such constraints serves as an indicator of the \textit{effective activation energy}, which arises from steric effects, and prevents the network from reorganizing locally to accommodate intermediate units forming over the course of the dissolution.
\end{abstract}

\maketitle

\section{Introduction}

Due to their abundance in Earth's crust, silicate minerals and glasses are the base components of numerous industrial materials including glasses, ceramics, and concrete. As such, understanding and predicting the kinetics of silicate dissolution in the presence of water has fundamental and industrial implications, with examples including bioactive glasses \cite{jones_review_2013}, nuclear waste confinement matrices \cite{harvey_model_1984, grambow_nuclear_2006} and the reaction of cement with water, the binding phase in concrete \cite{taylor_cement_1997}, all of which require accurate predictions of dissolution rates, in the context of microstructure evolution, phase stability and chemical durability.

Predicting the kinetics of dissolution of materials under given thermodynamic conditions is, a priori, a complex problem as it depends, amongst others, on the solvent chemistry \cite{cailleteau_insight_2008} and the geometry of the surface \cite{anbeek_surface_1992, brantley_kinetics_2008}. Another difficulty arises from the fact that the dissolution of minerals can feature leaching, or incongruency, that is, occuring in a nonstoichiometric fashion \cite{brantley_kinetics_2008}. However, this latter feature typically is true only during the initial stages of contact with water, with steady-state dissolution usually showing congruency \cite{brantley_kinetics_2008, lanford_hydration_1979}. Nevertheless, for a given solvent and thermodynamic condition, both the composition and structure of a material dictate its dissolution kinetics \cite{casey_control_1992, ohlin_dissolution_2010}.

Mere knowledge of the composition of a material is insufficient to predict its dissolution rate as, e.g., glassy silica dissolves between one and three orders of magnitude faster than crystalline $\alpha$-quartz \cite{brantley_kinetics_2008}, depending on the solvent pH. While it may be argued that glasses should indeed be less stable that their crystalline equivalents, such a trend is not systematic as, e.g., albite glass and crystal show comparable dissolution rates \cite{hamilton_dissolution_2000}. In fact, the dissolution of silicate glasses and minerals has recently been shown to occur by similar mechanisms and to be mostly controlled by short-range atomic order, rather than the long-range disorder \cite{hellmann_nanometre-scale_2015}. Indeed, the dissolution rates of silicate glasses are generally thought to decrease with the connectedness, or the degree of polymerization of the atomic network \cite{hamilton_effects_1997}, described by the ratio of terminating non-bridging oxygen (NBO) per tetrahedral unit (T) formed by silicon or aluminum atoms (NBO/T). However, using such a metric is far too restrictive as, e.g., silica, albite, jadeite, and nepheline glasses all feature a fully polymerized network (NBO/T = 0), but show vastly differing dissolution kinetics \cite{hamilton_dissolution_2001}. As such, given the disordered nature of glass, it is a challenge to identify a metric that is simple enough to enable practical predictions of dissolution rates, but that takes into account enough structural detail to provide realistic results.

\begin{figure}
\begin{center}
\includegraphics*[width=\linewidth, keepaspectratio=true, draft=\ddst]{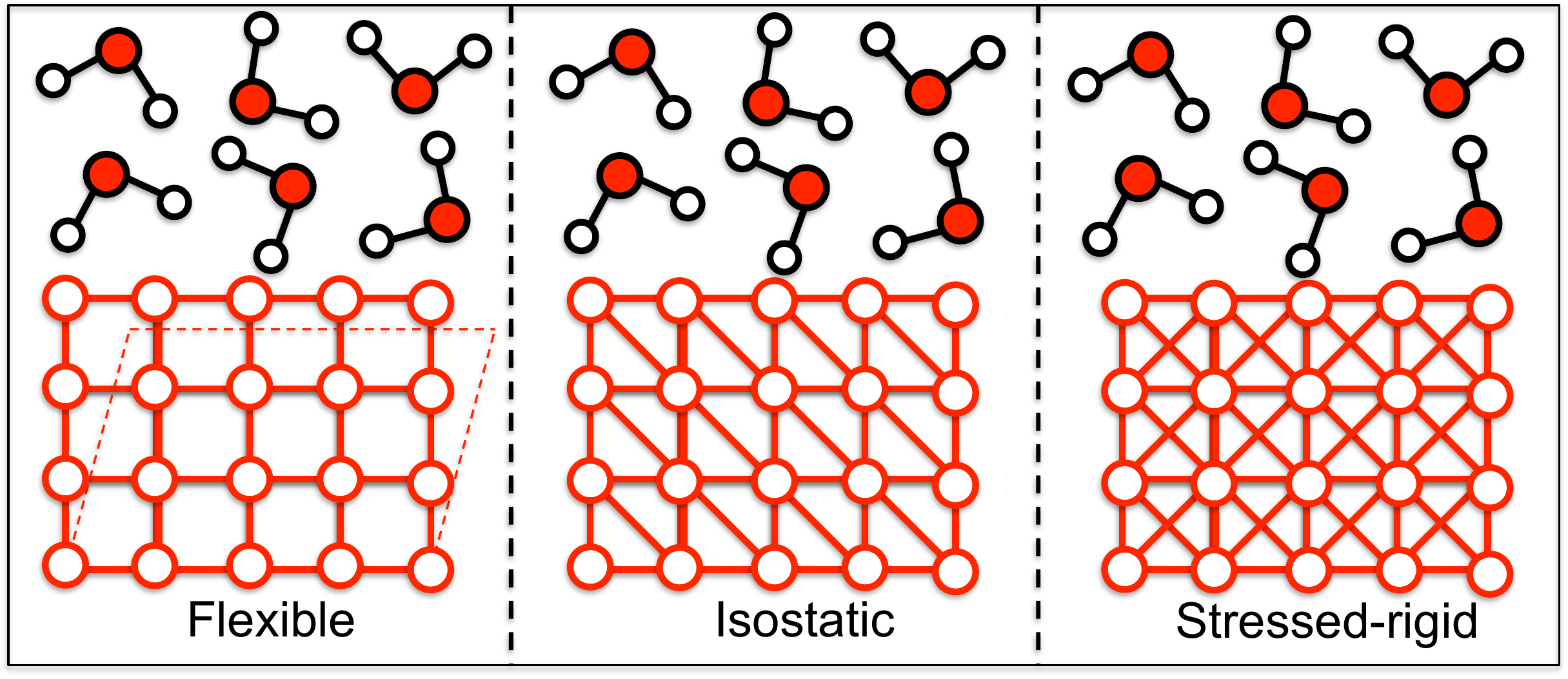}
\caption{\label{fig:rigidity} A sketch representing the three states of rigidity (flexible, isostatic, and stressed-rigid) of the atomic network of a solid in contact with water. The flexible state is the only one that features internal low-energy (floppy) modes of deformation.
}
\end{center}	
\end{figure}

By capturing the topology of atomic networks while filtering out chemical details that ultimately do not affect the macroscopic properties, topological constraint theory \cite{phillips_topology_1979, thorpe_continuous_1983, mauro_topological_2011} (TCT) is a promising tool that can elucidate such a metric. Within the TCT framework, atomic networks are described as mechanical trusses, in which the atoms experience topological constraints, as imposed by radial and angular chemical bonds. Hence, following Maxwell's stability criterion \cite{maxwell_l._1864}, an atomic network is described as flexible, stressed-rigid, or isostatic, if the number of topological constraints per atom ($n_{\rm c}$) is lower, higher, or equal to three, i.e., the number of degrees of freedom per atom in three dimensions (see Figure \ref{fig:rigidity}). In this paper, we establish that the dissolution rates of glassy and crystalline silicates in caustic environments are dictated by $n_{\rm c}$, which serves as an indicator of the \textit{effective activation energy} of the dissolution process, and, in turn, of the activation energies of solid state ion-diffusion and conduction. This suggests that, simply, the mechanical stability of the bulk atomic network controls the dissolution behavior, and the transport properties of silicate solids.

\section{Results}

\subsection{Dissolution rates}

\begin{figure}
\begin{center}
\includegraphics*[width=\linewidth, keepaspectratio=true, draft=\ddst]{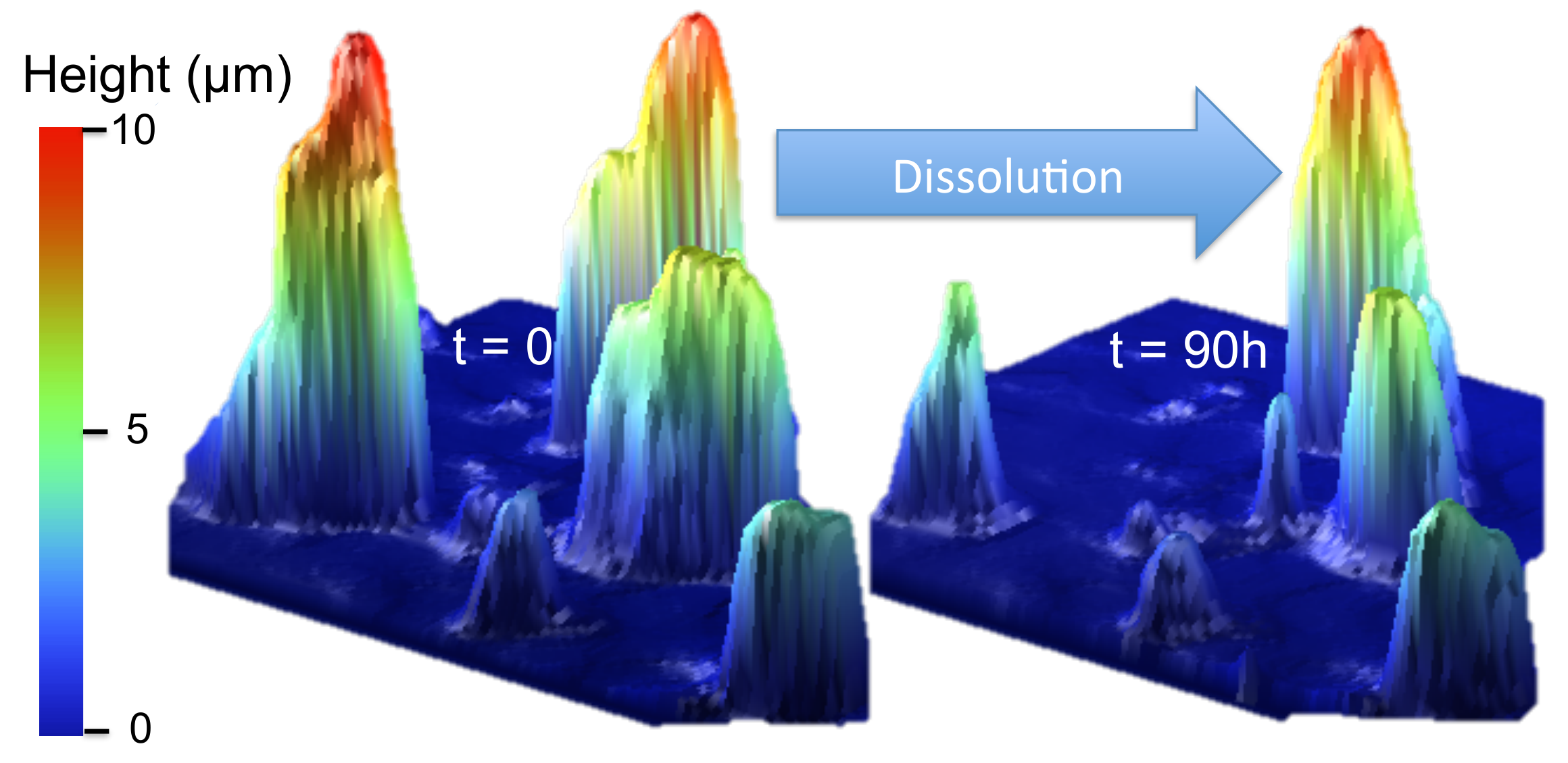}
\caption{\label{fig:vsi} An illustration of the dissolution of $\alpha$-quartz particulates, as visualized using vertical scanning interferometry (VSI), before and after 90h of solvent contact at pH 13 and 25 $^o$C. Dissolution is tracked by measuring the decrease in particle height over time.
}
\end{center}	
\end{figure}

The dissolution rates of glassy silica and $\alpha$-quartz were measured under isothermal conditions, and at different temperatures using vertical scanning interferometry \cite{kumar_vertical_2013} (VSI, see the Supplemental Material). This technique has been extensively applied to measure the dissolution rates of minerals of geological relevance \cite{lasaga_variation_2001, dove_mechanisms_2005}. By directly tracking the evolution of the surface topography in time, with sub-nanometer vertical resolution, VSI accesses the \textit{true} dissolution rate of a solid dissolving in a given solvent. Unlike dissolution assessments that are based on analysis of solution compositions, which may be affected due to aspects including metastable barrier formation, incongruency in dissolution or, ion adsorption, VSI analytics are not influenced by such complexities. Dissolution rates of the solids were quantified using a rain-drop procedure \cite{kumar_vertical_2013}, wherein both the solution pH and composition (i.e., the under-saturation level with respect to the dissolving solids) are kept constant over the course of a given experiment (see Figure \ref{fig:vsi}). In addition to our measurements, to assess the compositional dependence of dissolution kinetics, the dissolution rates of sodium trisilicate (Na$_2$O--3SiO$_2$), albite (Na$_2$O--Al$_2$O$_3$--6SiO$_2$), jadeite (Na$_2$O--Al$_2$O$_3$--4SiO$_2$), and nepheline (Na$_2$O--Al$_2$O$_3$--2SiO$_2$) glasses were sourced from the tabulations of Hamilton \textit{et al.} \cite{hamilton_dissolution_2000, hamilton_effects_1997, hamilton_dissolution_2001, hamilton_corrosion_1999}. Note that, for consistent comparisons between solids of different compositions, all the dissolution rates presented herein are normalized in terms of moles of O$_2$ dissolved per unit of surface and time.

\subsection{Relationship between composition and dissolution rates}

\begin{figure*}
\begin{center}
\includegraphics*[width=0.7\linewidth, keepaspectratio=true, draft=\ddst]{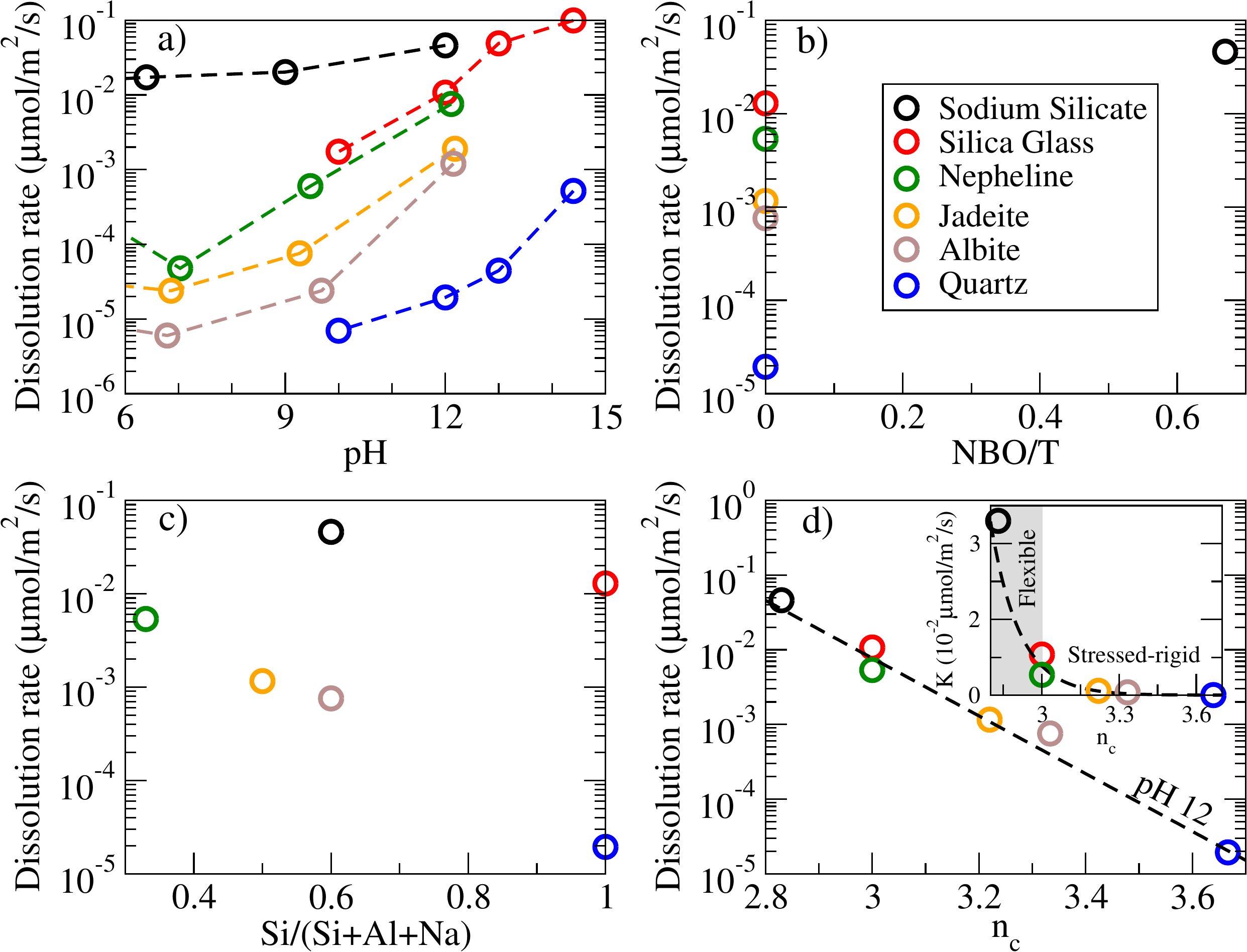}
\caption{\label{fig:rates} Influence of composition and structure on dissolution rate. (a) Room-temperature dissolution rates ($K$) of sodium silicate \cite{hamilton_corrosion_1999}, glassy silica, nepheline \cite{hamilton_dissolution_2001}, jadeite \cite{hamilton_dissolution_2001}, albite \cite{hamilton_dissolution_2001} and $\alpha$-quartz (normalized with respect to the amount of O$_2$ dissolved) versus solvent pH. (b) Dissolution rates at pH = 12 with respect to the ratio of non-bridging oxygen and Si or Al tetrahedra (NBO/T). (c) Same data as b), plotted with respect to the ratio of Si and cations atoms (Si/(Si+Al+Na)). (d) Same data as b), plotted with respect to the number of constraints per atom $n_{\rm c}$. The data are fitted by an Arrhenius-like law $K = K_0 \exp ( - n_{\rm c} E_0 \ RT )$, with $K_0$ = 1.4 10$^{11} {\rm \mu}$mol/m$^2$/s and $E_0$ = 25.5 kJ/mol (dashed line). The inset shows the same data as d), in linear scale. The grey area indicates the flexible domain ($n_{\rm c} <$ 3).
}
\end{center}	
\end{figure*}

Figure \ref{fig:rates}a shows dissolution rates for a range of solids in basic aqueous solvents. As expected, all the dissolution rates increase with pH, due to the increasing abundance of HO$^-$ species available for nucleophilic attacks of Si or Al tetrahedra \cite{hamilton_corrosion_1999}. Although the slope of the dissolution rates versus pH curve varies for each material, we note that the ranking of dissolution kinetics remains consistent across a wide pH range and scales as $\alpha$-quartz < albite < jadeite < nepheline < glassy silica < sodium silicate.

The structural origin of the dissolution rates can be elucidated by comparing dissolution rates for a representative pH, in this case pH 12. First, as shown in Figure \ref{fig:rates}b, relying on the degree of depolymerization, NBO/T, is insufficient. For example, if sodium silicate (NBO/T = 0.67) dissolves faster than glassy silica (NBO/T = 0), this metric does not distinguish the structures of nepheline, jadeite, albite and $\alpha$-quartz (NBO/T = 0) either, although their dissolution rates span three orders of magnitude. Second, one cannot rely on composition either. The ranking of the energies of Si--O (444 kJ/mol), Al--O (423 kJ/mol), and Na--O (83 kJ/mol) bonds suggests that the stability of a given silicate should increase with the fraction of silicon present, thereby decreasing the dissolution rate \cite{varshneya_fundamentals_1993}. Although, as shown in Figure \ref{fig:rates}c, this trend is true for nepheline, jadeite, and albite, it does not allow one to explain differences in dissolution rates between glassy and crystalline silica. In summary, none of the typical, and expected metrics are able to capture sufficient structural detail to permit quantitative estimation of the dissolution kinetics.

We now calculate the rigidity of the atomic networks of the different silicate solids to evaluate the relevance of this metric for dissolution. In a fully connected network, like, e.g., chalcogenide glasses, an atom of coordination number $r$ experiences $r$/2 radial bond-stretching (BS) constraints, and 2$r$ - 3 angular bond-bending (BB) constraints \cite{micoulaut_structure_2013}. This allows one to enumerate $n_{\rm c}$ = 3.67 constraints per atom for quartz \cite{bauchy_rigidity_2015} (stressed-rigid). In glassy silica, the Si--O--Si angle typically shows a much broader angular excursion \cite{yuan_si-o-si_2003}, so that this angle appears unconstrained. This results in $n_{\rm c}$ = 3.00 for glassy silica (isostatic), in agreement with the observation that silica is an excellent glass forming system \cite{bauchy_rigidity_2015}. The insertion of sodium cations typically substantially decreases the glass transition temperature of the glass and, thereby, restores the Si--O--Si BB constraint. In addition, each Na cation depolymerizes the base silica network by creating one NBO. The number of constraints per atom of (Na$_2$O)$_x$(SiO$_2$)$_{1-x}$ can then be calculated \cite{bauchy_atomic_2011} as being $n_{\rm c}$ = (11 - 10$x$)/3. This is in agreement with experimental evidence \cite{vaills_direct_2005} of a rigidity transition ($n_{\rm c}$ = 3) at $x$ = 0.20. Then, from a simplistic viewpoint, starting from sodium silicate, each added aluminum cation consumes one NBO. Na cations then act as charge compensators around Al$^{\rm IV}$ cations \cite{xiang_structure_2013} and are, therefore, not considered any further in the constraints enumeration. Note that the tetrahedral angular environment of Al cations is not as well defined as that of Si, so that the O--Al--O angles are considered unconstrained \cite{bauchy_structural_2014-1}. Consequently, the number of constraints per atom for (Na$_2$O)$_x$(Al$_2$O$_3$)$_x$(SiO$_2$)$_{1-2x}$ is given by $n_{\rm c}$ = (11 - 10$x$)/3. However, this enumeration under-estimates $n_{\rm c}$ as a small fraction of over-coordinated three-fold oxygen tricluster and five-fold aluminum species are found \cite{xiang_structure_2013, bauchy_structural_2014-1} . Hence, the rigidity of the aluminosilicate glasses was evaluated by a careful analysis of their structure, as predicted by the molecular dynamics (MD) simulations. This allows clear differentiation of intact from thermally broken BS (BB) constraints by computing the radial (angular) excursion of each neighbor, using a previously established methodology \cite{bauchy_atomic_2011, bauchy_topological_2012, bauchy_angular_2011} (see the Supplemental Material). As such, we find $n_{\rm c}$ = 3.33, 3.22, and 3.00 for albite, jadeite, and nepheline, respectively. Note that the latter value implies that nepheline features an isostatic network ($n_{\rm c}$ = 3), which is consistent with the fact that its dissolution rate is similar to that of glassy silica.

\subsection{Topology controls the kinetics of dissolution}

\begin{figure*}
\begin{center}
\includegraphics*[width=0.7\linewidth, keepaspectratio=true, draft=\ddst]{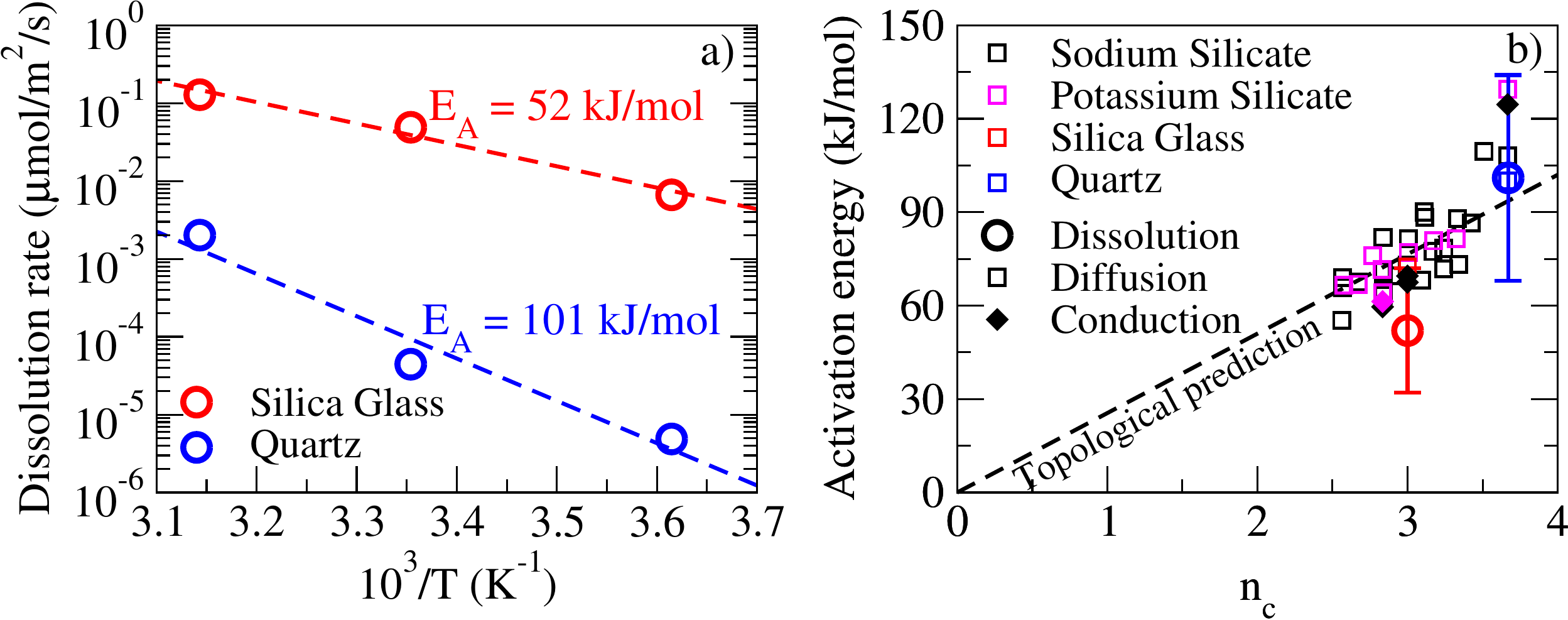}
\caption{\label{fig:energies} Activation energy of dissolution. (a) Dissolution rates of glassy silica and $\alpha$-quartz at pH 13 as a function of the inverse temperature 10$^3$/$T$. The data are fitted by an Arrhenius-like law $K = K_0 \exp ( - n_{\rm c} E_0 / RT )$, with $K_0$ = 6.4 10$^7 {\rm \mu}$mol/m$^2$/s and $E_{\rm A}$ = 52 kJ/mol for glassy silica, and $K_0$ = 1.6 10$^{14} {\rm \mu}$mol/m$^2$/s and $E_{\rm A}$ = 101 kJ/mol for $\alpha$-quartz (dashed lines). (b) The activation energy of the dissolution process for glassy silica and $\alpha$-quartz at pH 13, with respect to the number of constraints per atom $n_{\rm c}$. The data are compared with the activation energies of sodium silicate (self-diffusion \cite{frischat_ionic_1975} and conduction \cite{ngai_correlation_1989} of Na), potassium silicate (self-diffusion \cite{frischat_ionic_1975} and conduction \cite{ngai_correlation_1989} of K) fused silica (diffusion of water \cite{moulson_water_1961}), and quartz (diffusion of water \cite{shaffer_diffusion_1974}). The dashed line shows the effective activation energy predicted from the number of constraints per atom, following $E_{\rm A} = n_{\rm c}  E_0$, where $E_0$ is the activation energy required to break a given constraint per atom present in the atomic network.
}
\end{center}	
\end{figure*}

Figure \ref{fig:rates}d shows the dissolution rates $K$ as a function of the number of constraints per atom $n_{\rm c}$. As shown in the inset, $K$ significantly increases in the flexible domain ($n_{\rm c}$ < 3), in agreement with the idea that, due to the presence of internal low energy modes of deformation (i.e., floppy modes), a flexible network is less stable than its rigid counterparts, and should, therefore, dissolve faster. The observed evolution of $K$ with respect to the number of constraints $n_{\rm c}$ suggests an Arrhenius-like relationship, of the form:

\begin{equation}
  \label{eq:K}
  K = K_0 \exp \left( - \frac{n_{\rm c} E_0}{RT} \right)
\end{equation} where $K_0$ is a rate constant that depends on the solution phase chemistry, and yields the barrier-less dissolution rate of a completely depolymerized material (i.e., for which $n_{\rm c}$ = 0) and $E_0$ = 25.5 kJ/mol is an energy barrier that needs to be overcome to break a unit atomic constraint. This suggests that:

\begin{equation}
  \label{eq:Ea}
  E_{\rm A} ^{\rm eff} = n_{\rm c}  E_0
\end{equation} serves as an effective activation energy that controls the kinetics of the dissolution. To validate this hypothesis, we evaluated the low-temperature activation energy ($E_{\rm A}$) of dissolution of $\alpha$-quartz and glassy silica by measuring their dissolution rates at 3.5, 25, and 45 $^o$C at pH 13. We chose this higher pH to achieve appreciable dissolution of quartz at the lowest temperature. Note that these materials feature a different $n_{\rm c}$ (3.67 and 3.00 for quartz and glassy silica, respectively) but similar composition, which allows us to isolate the singular effect of atomic topology on the activation energy, with no compositional effects. As illustrated in Figure \ref{fig:energies}a, both materials show an Arrhenius-like evolution of their dissolution rate with temperature. As such, we find $E_{\rm A}$ = 101$\pm$33 kJ/mol and 52$\pm$20 kJ/mol for $\alpha$-quartz and glassy silica, respectively, in agreement with prior studies \cite{bunker_molecular_1994, zhang_molecular-level_2014} -- an expected outcome since $\alpha$-quartz shows a higher $n_{\rm c}$ than glassy silica (i.e., on account of its stressed-rigid nature). Interestingly, as shown in Figure \ref{fig:energies}b, these values are in fair agreement with the effective activation energy predicted simply from the knowledge of the number of constraints per atom ($E_{\rm A} ^{\rm eff} = n_{\rm c}  E_0$), which yields to $E_{\rm A} ^{\rm eff}$  = 93.6 kJ/mol and 76.5 kJ/mol for $\alpha$-quartz and glassy silica, respectively. This suggests the following atomistic picture: starting from $n_{\rm c}$ = 0, which would correspond to a fully depolymerized material, each new constraint per atom effectively reduce the dissolution kinetics by increasing the associated activation energy needed for bond rupture.

The activation energies for dissolution are compared to the activation energies of other thermally-activated processes, namely, for sodium silicate (self-diffusion \cite{frischat_ionic_1975} and conduction \cite{ngai_correlation_1989} of Na), potassium silicate (self-diffusion \cite{frischat_ionic_1975} and conduction \cite{ngai_correlation_1989} of K), fused silica (diffusion of water \cite{moulson_water_1961}), and quartz (diffusion of water \cite{shaffer_diffusion_1974}). First, we note that, interestingly, for a given material, the activation energies associated with dissolution, diffusion and conduction are consistent with each other, which suggests that these energy barriers all arise from a common atomistic origin -- which we identify as the rigidity of the network. Second, the activation energies all increase with the number of constraints per atom and are similar to the values predicted by equation \ref{eq:Ea}, that is, the effective activation energy induced by the number of constraints per atom.

\section{Discussion}

The question remains regarding the origin of the relationship that is found herein between the activation energy $E_{\rm A}$ and the number of topological constraints per atom $n_{\rm c}$. As such, we propose that $n_{\rm c}$ serves as an indicator of steric effects in the atomic network, which prevent reorganization and internal motion of the constituent species. The reaction mechanism of a silicate with water can take different forms: (1) hydration, where intact water molecules enter the glass network \cite{smets_role_1983}, (2) hydrolysis, where water reacts with Si--O bonds to form hydroxyl groups, and (3) ion-exchange, where network modifying cations, e.g. Na$^+$, are replaced by H$^+$ cations \cite{bunker_molecular_1994}. The three mechanisms can occur simultaneously, and influence process kinetics. For (1), the rate of water diffusion is primarily controlled by steric constraints, i.e., atomic packing, imposed by the network \cite{bunker_molecular_1994}. However, in typical silicates, the holes in the structure, as defined by the ring statistics, are too small to allow direct diffusion of intact water molecules into the network. Therefore, water would preferentially penetrate through hydrolysis and condensation reactions \cite{hamilton_corrosion_1999}. In mechanism (2), in basic solvents, Si tetrahedra are subjected to nucleophilic attack by HO$^-$ species, thereby forming five-fold coordinated Si intermediate units \cite{bunker_molecular_1994}. On the contrary, in acidic environments, inter-tetrahedral bridging oxygen atoms are subjected to electrophilic attack by protons (H$^+$) and form three-fold coordinated O intermediate units \cite{bunker_molecular_1994}. In these two cases, process kinetics depend on the ability of the network to accommodate such over-coordinated units by local rearrangements \cite{bunker_molecular_1994}. For (3), the rate-limiting step is suggested to be the penetration of the glass surface by water \cite{bunker_molecular_1994}. However, the three mechanisms are strongly coupled as the depolymerization of the silicate network due to (2) can open up the atomic network, e.g., due to ring opening thus enhancing water mobility for (1) and (3) to occur at an expedited rate.

In these three mechanisms, the kinetics of dissolution are limited by the ability of the atomic network to locally rearrange in order to: (a) accommodate intermediate over-coordinated atoms (Si$^{\rm V}$ or O$^{\rm III}$) or (b) enable the creation of pathways, or channels \cite{bauchy_pockets_2011}, by enlarging preexisting holes, thereby allowing a penetrating species (water molecule or H$^+$) to transfer between two sites. As such, the activation energy associated with such reorganizations takes the form of a strain energy that characterizes the ability of the network to locally resist, that is, to prevent the formation of energetic instabilities (departure from an equilibrium configuration) that are induced due to a dissolution process. Then, it becomes clear that the development of a strain energy, and its follow on dissipation, is linked to the number of constraints per atom $n_{\rm c}$, since each constraints acts as a little spring between the atoms. Therefore, $n_{\rm c}$ characterizes the steric effect imposed by the architecture of the atomic network, which prevents the accommodation of defects (i.e., over-coordinated atom or enlarged hole to create a channel). This picture is in line with result from density functional theory, which have shown that the activation energy of the hydrolysis of the inter-tetrahedra bridging oxygen increases with the extent of the steric constraints imposed by the network, that is, its connectivity and therefore, rigidity \cite{pelmenschikov_lattice_2000}.

\section{Conclusions}

Our results demonstrate that the dissolution rate of silicate-based glassy and crystalline solids in caustic solutions is correlated to the number of topological constraints per atom present in the network. Such atomic constraints are indicative of the effective activation energy that needs to be overcome for dissolution to occur. Moreover, $n_{\rm c}$ characterizes the ease with which the atomic network can reorganize to accommodate the intermediate (metastable) defects states that are formed over the course of the net dissolution process. The rate at which such reorganization can occur, appears to be the primary rate-controlling step during dissolution, that is, which controls its kinetics. This leads to the idea that dissolution kinetics, a response through to be controlled by surface properties, is paradoxically primarily controlled by the topology of the bulk architecture of the atomic network of the dissolving material. In addition, it is remarkable that a number of processes that share a common thermally activated origin, e.g., dissolution, diffusion and transport show similar trends in activation energy with respect to the number of topological constraints per atom. If, indeed, it is well known that diffusion and conductivity are related to each other via the Nernst-Einstein equation, it is likely that dissolution shares a similar basis, although perhaps better described by a Butler-Volmer type of formulation \cite{kristiansen_pressure_2011}. Nevertheless, our results suggest that dissolution, diffusion, and conduction could potentially be treated within the same atomistic, and thermodynamic framework -- enabling a priori estimation of such phenomena. This has important implications for better understanding and modeling processes ranging from cement hydration, to the chemical durability of glasses used to encapsulate radioactive wastes, to rate controls on silicate mineral dissolution that accompanies ocean acidification \cite{orr_anthropogenic_2005}.

\begin{acknowledgments}
The authors acknowledge full financial support for this research provided by: The U.S. Department of Transportation (U.S. DOT) through the Federal Highway Administration (DTFH61-13-H-00011), the National Science Foundation (CMMI: 1066583 and CAREER Award: 1235269), The Oak Ridge National Laboratory operated for the U.S. Department of Energy by UT-Battelle (LDRD Award $\#$ 4000132990), and the University of California, Los Angeles (UCLA). Access to computational resources was provisioned by the Physics of AmoRphous and Inorganic Solids Laboratory (PARISlab), the Laboratory for the Chemistry of Construction Materials (LC$^2$) and the Institute for Digital Research and Education (IDRE) at UCLA. This research was conducted in: Laboratory for the Chemistry of Construction Materials (LC$^2$) and Physics of AmoRphous and Inorganic Solids Laboratory (PARISlab) at UCLA. The authors gratefully acknowledge the support that has made these laboratories and their operations possible. The contents of this paper reflect the views and opinions of the authors, who are responsible for the accuracy of the datasets presented herein, and do not reflect the views and/or policies of the funding agencies, nor do the contents constitute a specification, standard or regulation.
\end{acknowledgments}

\end{document}